\documentclass{emulateapj}

\usepackage{subfigure,epsfig,url,bm,amsmath,cases,color,soul,longtable,orcidlink}

\begin{document}

\title{Evaluating an Embryo Origin for Detached TNOs within Full Kuiper Belt Formation Models}

\author{Nathan A. Kaib\orcidlink{0000-0001-5272-5888}}
\affiliation{Planetary Science Institute, 1700 E. Fort Lowell, Suite 106, Tucson, AZ 85719, USA}
\author{Chadwick A. Trujillo\orcidlink{0000-0001-9859-0894}}
\affiliation{Department of Astronomy and Planetary Science, Northern Arizona University, PO Box 6010, Flagstaff, AZ 86011, USA}
\author{Scott S. Sheppard}
\affiliation{Earth and Planets Laboratory, Carnegie Institution for Science, 5241 Broad Branch Road NW, Washington, DC 20015, USA;}

\begin{abstract} 
With perihelia well beyond Neptune, but semimajor axes and eccentricities indicative of substantial perturbation, the origins of detached trans-Neptunian objects (TNOs) remain a dynamical puzzle. In particular, detached TNOs with orbital inclinations below $\sim$25$^{\circ}$ are not easily generated from any known mechanism currently in the modern solar system. One notable hypothesis for the origins of detached TNOs is that a $\sim$Mars- to Earth-mass planetary embryo detached the perihelia of these objects from Neptune during the process of Kuiper belt formation before the embryo itself was ejected. We numerically model this scenario via simulations of Kuiper belt formation from a primordial planetesimal belt that is dispersed through the migration of the giant planets. In addition to $\sim$$10^5$ Kuiper belt objects, each of our simulations contains a hypothetical population of embryos in the primordial belt. We find that our embryos are unlikely to reach the high-perihelion, large semimajor axis orbit necessary to efficiently detach TNO perihelia from Neptune's influence. Moreover, embryos will typically take at least 100 Myrs to reach these unlikely orbits, at which point most of the primordial belt will have already been ejected by the planets, limiting the available population that can be detached. Finally, the TNOs that our embryos do detach consistently have a semimajor axis distribution that is more biased toward small values than observed detached TNOs have. Thus, we conclude that planetary embryos in the primordial Kuiper belt are not likely to have been the primary mechanism for the origin of detached TNOs. 
\end{abstract}


\section{Introduction}

The Kuiper belt is generally thought to be the remnant of a more massive primordial belt of planetesimals that were dynamically dispersed through gravitational interactions with the newly formed giant planets, in particular Neptune \citep[e.g.][]{mal95, lev08, nes18}. Consequently, most trans-Neptunian objects (TNOs) reside on orbits with significant orbital eccentricities and/or inclinations \citep[e.g.][]{glad08, bann18}, with the exception of the cold classical belt subpopulation, which is thought to have formed in-situ \citep{brown01, dor02, dawclay12, gladvolk21}. Kuiper belt formation models have demonstrated that most of these excited Kuiper belt orbits can be replicated via direct or resonant gravitational interactions with Neptune \citep{mal95, lev08, nes15b, volkmal19}.

However, not all aspects of the Kuiper belt are readily replicated with such models. In particular the Kuiper belt contains a subpopulation of TNOs whose perihelia are far from Neptune ($q\gtrsim40$ au) but whose eccentricities and semimajor axes are quite large ($e\gtrsim0.25$; $a>50$ au). These objects' eccentricities are similar to those expected if they had undergone close scattering encounters with the giant planets, but their large perihelia prevent such encounters from occurring on their modern orbits \citep{glad02}. If TNOs are trapped in resonance with Neptune, Kozai-Lidov cycles can pull their perihelia away from Neptune while conserving the vertical component of their orbital angular momentum \citep{koz62, lid62, gom03}. This increased pericenter weakens the coupling with Neptune, and objects can exit mean motion resonance during this high-inclination phase of Kozai cycles \citep{gom05}. If Neptune is still migrating away from the Sun during this process \citep{fernip84}, these high-perihelion bodies can ultimately be left far from resonances as the resonance locations continue to migrate with Neptune's semimajor axis \citep{kaibshep16}. 

Yet, not all high-perihelion TNO orbits can be explained with Kozai-Lidov cycling. One feature of TNO orbits generated with this mechanism is that they are typically found at high inclinations ($i\gtrsim25^{\circ}$) while at $q>40$ au \citep{shep16, kaibshep16, andkaib21}, but a significant number of known $q>40$ au TNOs are found at inclinations well under 25$^{\circ}$ \citep{beau23}. Consequently, a number of other origin mechanisms have been proposed. These include a close stellar encounter while the Sun inhabited its stellar birth cluster \citep{fern97, morblev04, bras06, nes23}, a distant still-undetected planet \citep{trujshep14, batbrown16, lykito23}, and one or more $\sim$Mars--Earth-mass planetary embryos embedded in the primordial Kuiper belt \citep{gladchan06, lykmuk08}. 

For the remainder of this paper, we loosely refer to excited TNOs with $q>40$ au objects as detached TNOs. Strictly speaking, many of these bodies are not fully dynamically decoupled from the giant planets, as they can still weakly scatter off the giant planets \citep[e.g.][]{bann17}, and the critical perihelion beyond which significant semimajor axis evolution ceases is actually a function of semimajor axis and inclination \citep{fern81, hadtre24}. Moreover, a significant fraction occupy high-order mean motion resonances (MMR) with Neptune \citep{volkvan24}. Thus, the dynamical behavior of this $q>40$ au population is diverse and not always fully decoupled from the planets. Nonetheless, in practice, known dynamical mechanisms within traditional Kuiper belt formation models struggle to place TNOs onto eccentric orbits with $q\gtrsim40$ au without also driving them to high inclination, and we adopt the ``detached'' nomenclature for simplicity's sake. 

In the detached TNO origin scenarios involving a stellar passage or a distant planet, the perturbations to detached TNOs are applied from a source that is well outside the solar systems's giant planet region, but this is not true for planetary embryos. In this last case, each embryo presumably forms near the giant planets before being scattered to an eccentric, large semimajor axis orbit, at which point its gravitational influence begins detaching the perihelia other TNOs from the giant planets \citep{gladchan06, siltre18}. Recent works have explored the effects of specific embryo orbits. \citet{huang22} argued that the observed Kuiper belt population on semimajor axes between 50--100 au is well-matched if an Earth-mass embryo occupies an orbit of $q\simeq40$ au and $a\simeq300$ au for $\sim$100 Myrs during the process of Kuiper belt formation. A somewhat similar embryo orbit of $q\simeq32$ au and $a\simeq400$ au was proposed in \citet{huangglad24}, as they found that the embryo detached TNO perihelia with Sedna-like semimajor axes (200--1000 au) and aligned them apsidally. In either case, it is assumed that energy kicks from the giant planets ultimately eject the embryo after it has built a population of detached TNOs, but it is also possible for it to survive to the modern epoch on a high-perihelion, large-semimajor axis orbit \citep{siltre18}. (We should note that the surviving embryo orbits predicted in \citet{siltre18} have substantially smaller perihelia and semimajor axes than that proposed for an additional super-Earth planet \citep{brownbat16}.)

In this origin scenario for the detached TNOs, the embryo actually participates in the process of Kuiper belt formation, and its evolution can in turn be significantly influenced by the combined gravity of smaller Kuiper belt objects. To date, works modeling the evolution of TNOs under the influence of one or more planetary embryos typically have either treated other Kuiper belt objects as massless test particles \citep{huangglad24} or used a statistically small number of bodies unable to resolve the details of the detached TNO population \citep{siltre18}. However, GPU-accelerated N-body integrators, such as GENGA \citep{grimmstad14}, offer an opportunity to better model the gravity of the Kuiper belt during during its formation while employing large enough numbers of particles to yield a modern belt populous enough to statistically characterize its orbital architecture \citep{kaib24}. If such simulations are begun with one or more planetary embryos embedded in the primordial belt, they can better characterize how embryo orbits evolve during Kuiper belt formation and provide new predictions on the population of detached TNOs that should result from the embryo's presence. 

In the following work, we perform a suite of GPU-accelerated simulations that model Kuiper belt formation with the inclusion of hypothetical populations of planetary embryos. In Section \ref{sec:meth} we describe the details of our simulation parameters and initial conditions. In Section \ref{sec:res}, we discuss the outcomes of our simulations and compare them with the observed distribution of detached TNOs in the actual solar system. Finally, in Section \ref{sec:con}, we conclude that planetary embryos are unlikely to attain orbits that can efficiently generate detached TNOs and even within these orbits they are unable to replicate the semimajor axis distribution of the observed detached TNO population.

\section{Dynamical Simulation Methods}\label{sec:meth}

To understand how the presence of planetary mass embryos can detach TNOs during Kuiper belt formation, we perform a set of Kuiper belt formation simulations that are very similar to those presented in \citet{kaib24}. These simulations contain 89,000 bodies whose individual masses are $\sim$0.09 M$_{\rm Pluto}$ and another 1,000 bodies whose individual masses are 1 M$_{\rm Pluto}$. The gravity of the Pluto-mass bodies is fully included in our simulations, whereas the sub-Plutos are treated as `semi-active,' in that they do not feel one another's gravity, but more massive bodies feel their gravity and vice versa. The initial orbits of our Pluto and sub-Pluto bodies are nearly circular and coplanar ($e<.01$; $i<1^{\circ}$), and their semimajor axes are randomly selected from a uniform distribution between 23.4 and 30 au. Each disk of Plutos and sub-Plutos is exterior to a resonant chain of giant planets (Jupiter, Saturn, and three 15.8 M$_{\oplus}$ ice giants). These planets begin in a 3:2, 3:2, 2:1, 3:2 resonant chain with the outermost semimajor axis initially at 20.4 au \citep{nesmorb12, kaib24}. While other early giant planet evolution scenarios have been proposed \citep[e.g.][]{clem21b, clem21a}, we choose this one as a test case given its prior study. 

In addition to the aforementioned bodies, a small number of super-Pluto-mass embryos are also implanted into the disk of each system (with randomly selected semimajor axes). These embryos' masses range from an Earth mass down to a lunar mass, and each simulation considers a set of equal-mass embryos. The number of embryos ranges from 3 (for our largest masses) up to 100 (for our smallest masses). We also include a set of control systems with no embryos. A summary of our simulated systems are listed in Table \ref{tab:IC}. 

\begin{table}
\begin{tabular}{c c c c}
\textbf{Simulation Initial Conditions} & & & \\
\end{tabular}
\centering
\begin{tabular}{c c c c c c c}
\hline
Sim & $M_{disk}$ & $N_{Pluto}$ & $N_{super}$ & $M_{super}$ & $a_{min}$ & $a_{max}$ \\
 & ($M_{\oplus}$) & & &  ($M_{\oplus}$) & (au) & (au) \\
\hline
3E & 23 & 1000 & 3 & 1.0 & 23.4 & 30 \\
20M & 22.14 & 1000 & 20 & 0.107 & 23.4 & 30 \\
5M & 20.54 & 1000 & 5 & 0.107 & 23.4 & 30 \\
2M & 20.21 & 1000 & 2 & 0.107 & 23.4 & 30 \\
100L & 21.23 & 1000 & 100 & 0.012 & 23.4 & 30 \\
30L & 20.37 & 1000 & 30 & 0.012 & 23.4 & 30 \\
10L & 20.12 & 1000 & 10 & 0.012 & 23.4 & 30 \\
Control & 20 & 1000 & 0 & n/a & 23.4 & 30 \\
\hline
\end{tabular}
\caption{Table of simulation initial conditions. From left to right, the columns are as follows: (1) simulation name, (2) total mass of primordial belt, (3) initial number of Pluto-mass bodies, (4) initial number of super-Pluto bodies, (5) mass of super-Pluto body used, (6) minimum semimajor axis of primordial belt bodies, (7) maximum semimajor axis of primordial belt bodies.}
\label{tab:IC}
\end{table}

Each set of initial conditions from Table \ref{tab:IC} is generated twice, with a uniquely drawn planetesimal disk in each instance. These systems are then integrated until the onset of a planetary orbital instability. Because any individual instability is unlikely to yield a good analog to our outer planets' orbital architecture \citep{nesmorb12, kaib24}, after a system reaches instability, it is repeated from the point of instability (the first output after a close planet-planet encounter) with each component of the innermost ice giant's cartesian position by a random amount between $\pm5\times10^{-10}$ au. This is done until each initial realization yields three different post-instability states that we deem to be ``solar system-like,'' in that one ice giant is ejected, and two survive on orbits exterior to Jupiter and Saturn, which cross their 2:1 MMR but remain below a period ratio of 2.8. This process yields six different post-instability states for each row in Table \ref{tab:IC}. Each system is integrated (pre- and post-instability combined) for a total of 4 Gyrs. 

Our simulation pipeline yields 48 different final Kuiper belt orbital architectures. However, subsequent analysis reveals that the Neptunes in some of our systems experience eccentricity excursions that are extreme compared to our solar system's modern state and its inferred history \citep{dawclay12, nes21}. This is an issue for our purposes because we focus our work on the population of TNOs that embryos detach, but a very eccentric phase of Neptune's orbital evolution can also detach TNOs and obscure the role of embryos. Thus, we post facto choose to exclude systems in which Neptune's eccentricity maximum exceeds 0.2 and/or systems where Neptune's final mean eccentricity is above 0.1. This lowers our number of analyzable systems from 48 down to 38. Amongst these 38 systems, the median final eccentricity of Neptune is 0.006, and only 4 of 38 systems have final Neptunian eccentricities over 0.03. In these 38 systems, we do not observe any strong correlation between final Kuiper belt properties and Neptune's maximum or final eccentricity.

Moreover, as in \citet{kaib24}, most (29 out of 38) of our simulated Neptunes also finish with an orbital period over twice that of Uranus, and the median value is 2.13. For Neptune-Uranus period ratios between 1.96--2.02, the distant resonances with Neptune are known to be significantly weaker than in the real solar system \citep{grahamvolk24}. This resonance-weakening effect is potentially important to the process of embryo-detaching, as many TNOs detached in prior simulations first spend time trapped in distant Neptunian resonances before becoming detached \citep{huang22}. However, only two of our 38 simulated systems finish with Neptune-Uranus period ratio between 1.96 and 2.02. (Their exact values are 1.964 and 1.970.) Of the 29 systems that finish with a Neptune-Uranus period ratio over 2.02, 20 of them evolve beyond a period ratio of 2.02 within 20 Myrs of the planetary instability. Thus, 27 of our 38 systems spend all or nearly all of their post-instability state outside of the 1.96--2.02 range of period ratios, and we expect capture into the distant Neptunian resonances in the large majority of our simulations to be at least as strong as those of the real solar system.

\subsection{Ejection Distance}\label{sec:eject}

Particles are removed from our simulations upon collision with another body, including the Sun, which we inflate to a radius of 0.5 au to stop the integration of extremely low perihelion passages \citep{levdun00}. The original intent of our simulations focused on more proximate resonant populations within the Kuiper belt. As a result, particles are also removed when they exceed a heliocentric distance of 1000 au, as our simulations do not include perturbations from the local Galactic environment, which become dynamically significant beyond this distance \citep{kaib11}. To guard against our ejection criterion from distorting our orbital distributions, we only consider simulated and observed TNOs with semimajor axes below 400 au throughout this paper. In principle, this 1000 au ejection distance could still distort our distribution of detached TNOs, since their semimajor axes typically undergo a random walk before their detachment. If TNOs often attain semimajor axes well above their final detachment semimajor axis before detaching, this could artificially remove some TNOs that would ultimately become detached.

To gauge whether this is a significant issue, we can study our simulations' population of actively scattering TNOs and see how well they replicate the observed scattering population. These bodies are continually randomly walking in semimajor axis while they gravitationally scatter off the giant planets (typically Neptune). To select scattering particles from our simulations, we measure their absolute change in semimajor axis over 10 Myrs of integration. If the change is greater than 1.5 au, they are classified as scattering \citep{glad08, law18}. We select scattering bodies in this way from the last 100 Myrs of each of our six Control simulations, sampling each system every 10 Myrs. These scattering particle orbits are then run through a survey simulator simulating the detection probabilities of the Canada-France Ecliptic Plane Survey \citep[CFEPS; ][]{petit11}, its high-latitude extension \citep[HiLat; ][]{petit17}, the survey of \citet{alex16}, and the Outer Solar System Origins Survey \citep[OSSOS; ][]{bann18}, which we collectively refer to as OSSOS+. When sampling our scattering particles, we randomly draw absolute magnitudes from the ``knee'' size distribution of \citet{law18}, which assumes a $\alpha=0.9$ power-law bright-end size frequency distribution (SFD) that transitions to a $\alpha=0.4$ faint-end SFD at $H=7.7$. In this manner, we build a sample of 1000 synthetic OSSOS+ detections from our Control simulations. 

\begin{figure}
\centering
\includegraphics[scale=0.4]{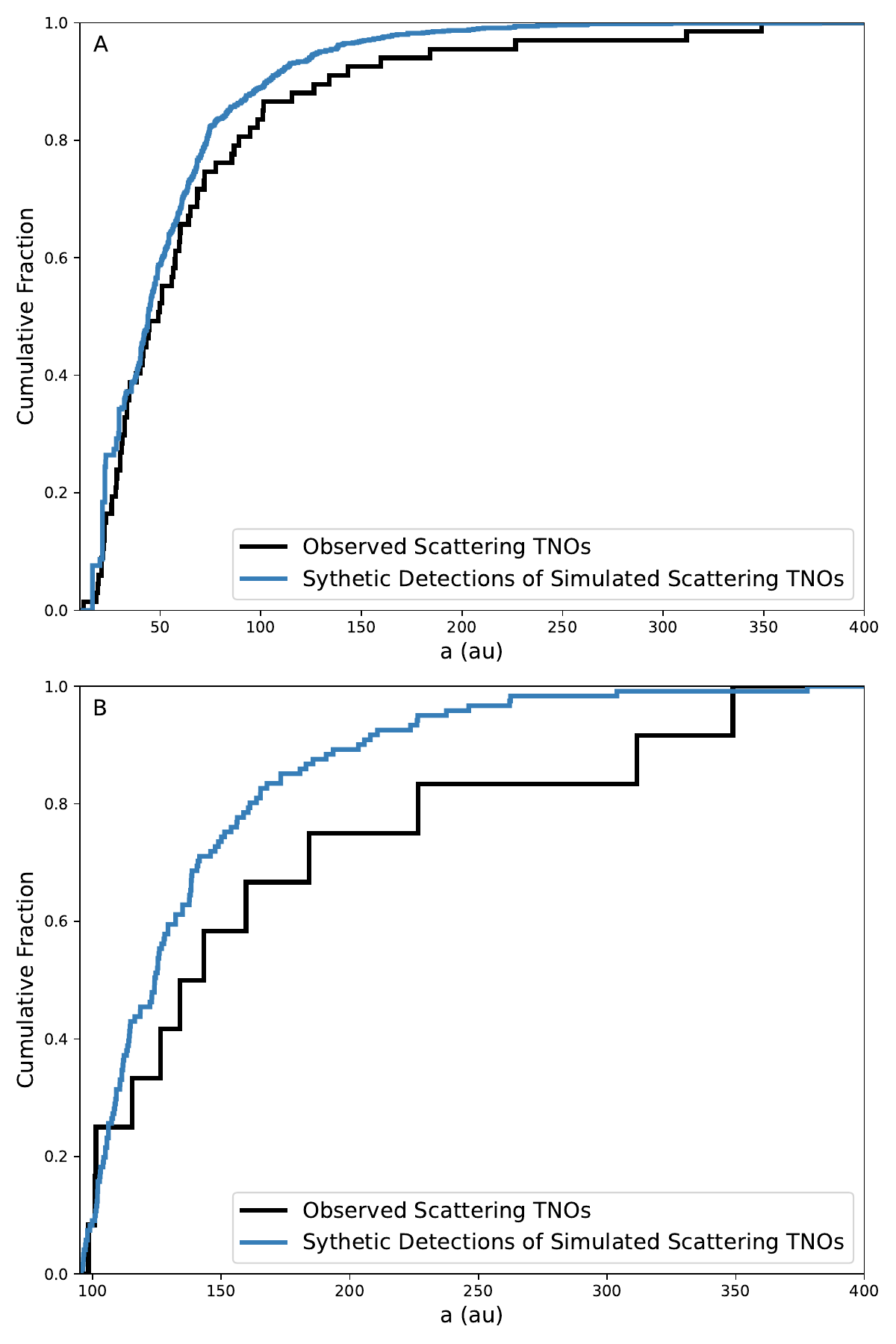}
\caption{{\bf A:} Cumulative semimajor axis distribution of OSSOS+ scattering TNOs ({\it black}) and simulated detections of scattering TNOs from our control simulations ({\it blue}). {\bf B:} The same as Panel A, except only TNOs with semimajor axes over 95 au are plotted. In both panels, the synthetic distribution is statistically consistent with the observed one.}
\label{fig:surfdenfit}
\end{figure}

In Figure \ref{fig:surfdenfit}A, the semimajor axis distribution of our synthetic scattering detections is shown along with that of the OSSOS+ catalog's 67 scattering objects with $a<400$ au. Here we see by-eye that there is modestly fewer large semimajor axis orbits in our simulations compared to the real solar system. However, the statistical significance of this deficit is not strong. A Kolmogorov-Smirnov (K-S) test comparing the two distributions returns a $p$-value of 0.35, while an Anderson-Darling (A-D) test returns a $p$-value of 0.16. This indicates we cannot confidently reject the null hypothesis that the two samples have the same underlying distribution. As we will see in subsequent sections, our analysis of detached TNOs will focus on those with semimajor axes over 95 au. Thus, in Figure \ref{fig:surfdenfit}B, we next compare the synthetic and observed distributions of only scattering bodies with $a>95$ au. We again see by-eye an apparent deficit of the largest semimajor axis orbits. As with the full semimajor axis distribution, though, the difference is not statistically significant. K-S and A-D tests return $p$-values of 0.59 and 0.19, respectively. Thus, we conclude that our ejection distance does not significantly distort our simulations' semimajor axis distribution inside $\sim$400 au.

\section{Results and Discussion}\label{sec:res}

\begin{table}
\begin{center}
\begin{tabular}{c c c c}
\hline

\hline 
Sim & $N_{\rm KB}$ & $N_{95<a<400}$ & a$_{90\%}$ \\ 
 & & & (au) \\
\hline 
\bf 3E\_1 & \bf 502 & \bf 61 & \bf 167 \\
\bf 3E\_2 & \bf 383 & \bf 75 & \bf 211 \\ 
\bf 3E\_3 & \bf 175 & \bf 2 & \bf  110 \\
\bf 3E\_4 & \bf 1043 & \bf 398 & \bf  213 \\
3E\_5 & 586 & 0 &  - \\
3E\_6 & 1405 & 2 &  - \\
20M\_1 & 488 & 1 &  - \\
\bf 20M\_3 & \bf 375 & \bf 24 & \bf  133 \\
\bf 20M\_4 & \bf 348 & \bf 9 & \bf  124 \\
\bf 20M\_5 & \bf 360 & \bf 3 & \bf  136 \\
\bf 20M\_6 & \bf 284 & \bf 1 & \bf  121 \\
5M\_1 & 282 & 0 &  - \\
\bf 5M\_3 & \bf 260 & \bf 1 & \bf  155 \\
5M\_5 & 254 & 1 &  - \\
5M\_6 & 520 & 1 &  - \\
2M\_1 & 293 & 0 &  - \\
2M\_2 & 543 & 1 &  - \\
\bf 2M\_3 & \bf 321 & \bf 6 & \bf 143 \\
2M\_6 & 265 & 0 &  - \\
100L\_1 & 329 & 0 &  - \\
100L\_2 & 381 & 1 &  - \\
100L\_4 & 131 & 0 &  - \\
100L\_5 & 412 & 0 &  - \\
\bf 100L\_6 & \bf 464 & \bf 2 & \bf 108 \\
30L\_1 & 251 & 0 &  - \\
30L\_3 & 458 & 0 &  - \\
30L\_5 & 453 & 0 &  - \\
10L\_1 & 342 & 0 &  - \\
10L\_2 & 239 & 0 &  - \\
10L\_3 & 720 & 0 &  - \\
10L\_4 & 471 & 0 &  - \\
10L\_6 & 544 & 0 &  - \\
Control\_1 & 125 & 0 &  - \\
Control\_2 & 834 & 0 &  - \\
Control\_3 & 261 & 0 &  - \\
Control\_4 & 414 & 0 &  - \\
Control\_5 & 94 & 0 &  - \\
Control\_6 & 147 & 0 &  - \\
\hline
\label{tab:fin}\\
\end{tabular}
\end{center}
\caption{Overview of simulation results. Left to right, the columns are (1) simulation name, (2) final number of particles remaining in simulation, (3) final number of low-inclination ($i<25^{\circ}$) detached ($q>40$ au) particles with semimajor axes between 95--400 au, (4) the 90th percentile semimajor axis of the observationally biased, time-sampled (once per 100 Myrs for final Gyr) distribution of particles with $i<25^{\circ}$, $q>40$ au, and $95<a<400$ au. Bolded rows correspond to simulations that possess 10 or more time-sampled orbits particles with $i<25^{\circ}$, $q>40$ au, and $95<a<400$ au.}
\end{table}

Because each of our Kuiper belt formation simulations undergoes a different planetary instability, Neptune's final semimajor axis varies by roughy $\pm1$ au across our simulations \citep{kaib24}. Thus, in each simulation, we rescale our particle's final semimajor axes according to Neptune's semimajor axis

\begin{equation}
a_{scaled} = 30.1 ~{\rm au} \times \frac{a_{sim}}{a_{Neptune,sim}}
\end{equation}

In this manner, major Neptunian mean motion resonances, which can detach the perihelia of distant TNOs, are located at the same semimajor axes across different simulations \citep{gom04, kaibshep16}. This rescaling of semimajor axis also results in the rescaling of the perihelia of simulated particles. 

\begin{figure}
\centering
\includegraphics[scale=0.4]{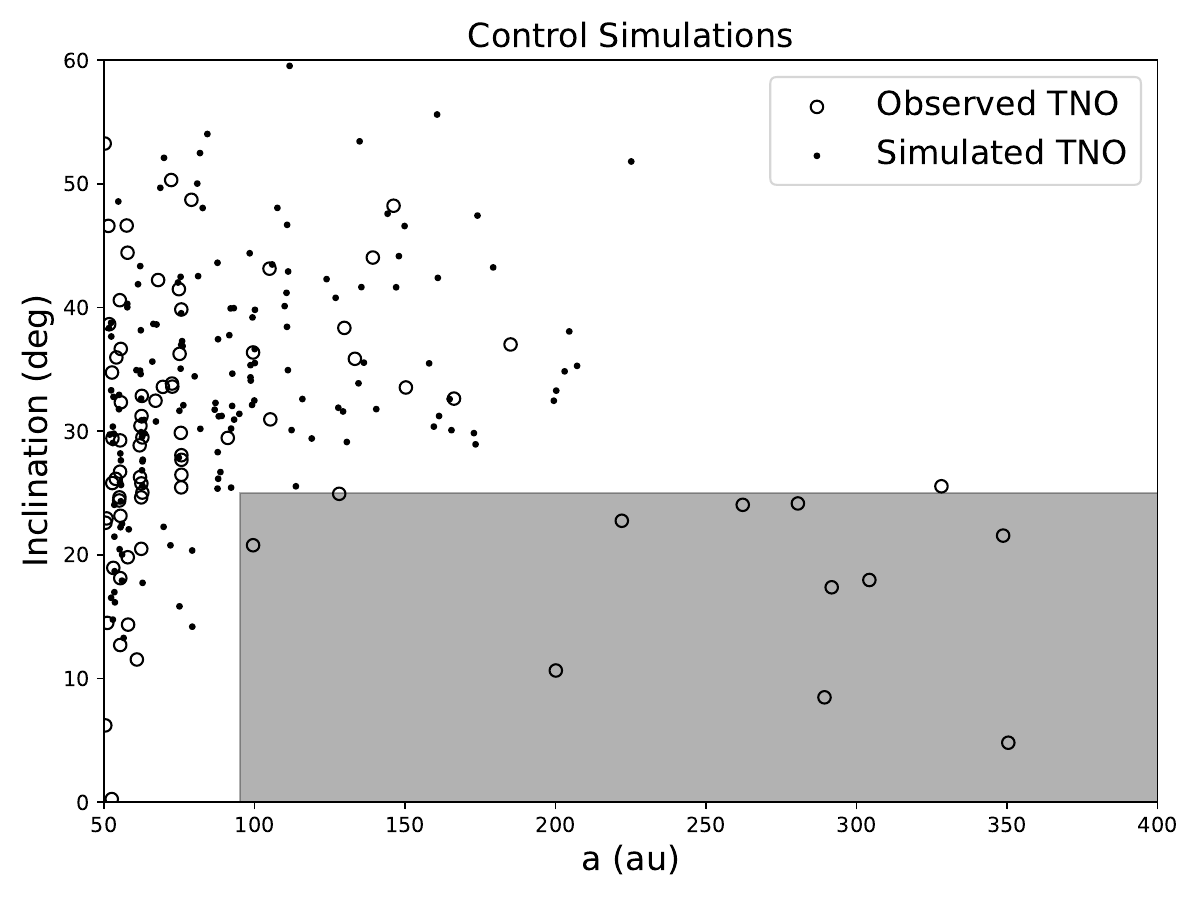}
\caption{Barycentric orbital inclinations vs semimajor axes of detached ($q>40$ au) TNOs. Open symbols are the orbits of observed TNOs, and filled symbols are the co-added final orbits of all detached ($q>40$ au) particles in the six Control simulations. Shaded region marks orbits with $95<a<400$ au and $i<25^{\circ}$, where TNOs are known to exist but control simulations produce no objects.}
\label{fig:lowincpop}
\end{figure}

We begin our simulation analysis with our six control simulations that feature no primordial belt bodies more massive than Pluto. In Figure \ref{fig:lowincpop}, we co-add all of our control simulations and plot the final orbital inclinations against the final orbital semimajor axes of all particles whose perihelia are beyond 40 au. In general, we see that inclinations over $\sim$$25^{\circ}$ are more common than lower inclinations. This is due to Kozai cycles acting on particles within Neptunian mean motion resonances. As the perihelia of such bodies are raised, so are the inclinations. This bias against low inclinations among detached ($q>40$ au) orbits becomes stronger as semimajor axis increases. Beyond semimajor axes of $\sim$80 au, no simulated particles exist with inclinations under $25^{\circ}$. Meanwhile, many simulated particles are seen with $i>25^{\circ}$ out to semimajor axes of 200--250 au, consistent with other works studying the decoupling of TNO perihelia within Neptunian MMRs \citep[e.g.][]{brasschwamb15, grahamvolk24}. 

In Figure \ref{fig:lowincpop} we also plot the orbits of all observed detached TNOs with inclination and perihelion uncertainties below $1^{\circ}$ and 1 au, respectively. (These uncertainties are of order 5--10
\% of the oscillation amplitudes that Kozai cycles impose on typical TNOs \citep{kaibshep16}.) By eye, we see that the observed TNO orbits and the simulated particles largely overlap with one another between semimajor axes of $\sim$50--80 au. However, the situation dramatically changes beyond $\sim$100 au. In this region of orbital space we see 11 observed TNOs with inclinations under 25$^{\circ}$, even though our control simulations appear incapable of populating this area. Since most TNO discovery surveys target the ecliptic, most have a strong bias toward detecting low inclination TNOs. However, surveys typically correct for this bias, and it is clear from decades of surveys that the TNOs contain a substantial number of high-inclination objects that are frequently detected even near the ecliptic \citep[e.g.][]{petit11, bann18, bern22}. The very existence of the observed low-inclination population in the shaded region of Figure \ref{fig:lowincpop} and its complete absence in our control simulations indicates the control simulations are missing one or more important processes. For reference, we list the names and orbits of these 11 observed TNOs in Table 3. (We should note that our $a<400$ au upper bound means this paper's analyses exclude nine other observed low-inclination detached TNOs whose orbital distances extend to or beyond $\sim$1000 au. These nine include well-studied objects such as Sedna, 2013 SY$_{99}$, and 2015 TG$_{387}$ \citep{brown04, bann17, shep19}. Again, this is done to ensure our simulations' ejection criteria do not impact our comparisons with the observed solar system.)

\begin{table}
\begin{center}
\begin{tabular}{c c c c}
\hline
Name & Semimajor Axis & Pericenter & Inclination\\
 & (au) & (au) & ($^{\circ}$)\\
\hline
2008 ST$_{291}$ & 99.47 & 42.45 & 20.78\\
2014 QS$_{562}$ & 128.09 & 40.35 & 24.94\\
2013 UT$_{15}$ & 200.06 & 43.93 & 10.65\\
2000 CR$_{105}$ & 221.98 & 44.12 & 22.76\\
2012 VP$_{113}$ & 262.16 & 80.52 & 24.05\\
2014 WB$_{556}$ & 280.48 & 42.7 & 24.16\\
2018 VM$_{35}$ & 289.32 & 44.54 & 8.48\\
2013 FT$_{28}$ & 291.71 & 43.5 & 17.38\\
2014 SR$_{349}$ & 304.23 & 47.45 & 17.97\\
2010 GB$_{174}$ & 348.67 & 48.59 & 21.56\\
2016 SD$_{106}$ & 350.39 & 42.71 & 4.81\\
\hline
\label{tab:obs}
\end{tabular}
\end{center}
\caption{List of known detached ($q>40$ au) TNOs with inclinations under 25$^{\circ}$ and semimajor axes between 95--400 au. Left to right, the columns list (1) object name, (2) semimajor axis, (3) pericenter, and (4) inclination. All orbital elements are barycentric.}
\end{table}

\begin{figure}
\centering
\includegraphics[scale=0.4]{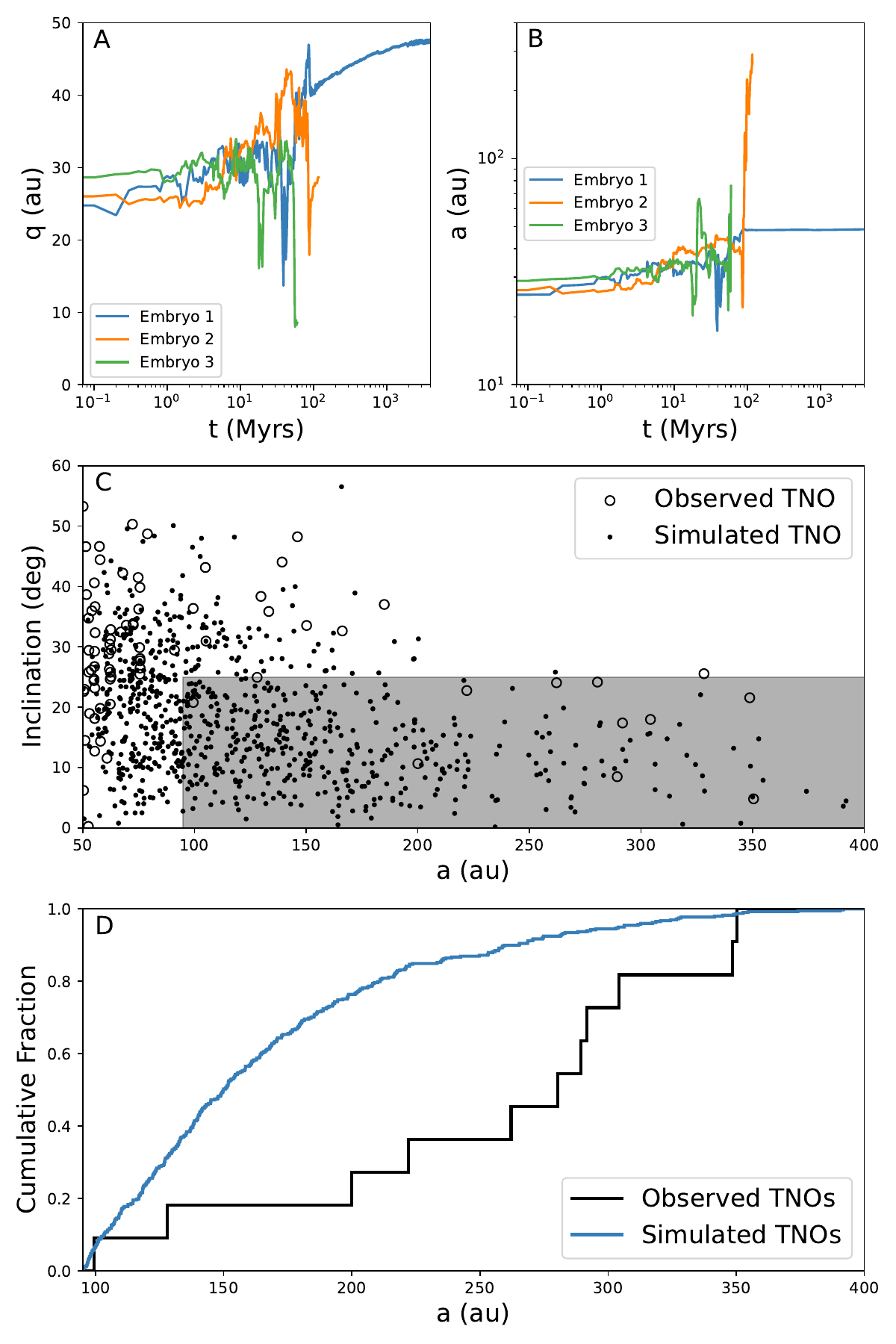}
\caption{{\bf A:} Time evolution of the pericenters of the three 1 M$_{\oplus}$ embryos in the 3E\_4 simulation. Each embryo is assigned a unique color. {\bf B:} Time evolution of the semimajor axes of the three 1 M$_{\oplus}$ embryos in the 3E\_4 simulation. Each embryo is assigned a unique color. {\bf C: } Barycentric orbital inclinations vs semimajor axes of detached ($q>40$ au) TNOs. Open symbols are the orbits of observed TNOs, and filled symbols are the final orbits of all detached particles in the 3E\_4 simulation. Shaded region marks orbits with $95<a<400$ au and $i<25^{\circ}$. {\bf D:} Cumulative semimajor axis distribution of TNOs falling in the shaded region of Panel C. Distributions are shown for observed TNOs ({\it black}) and TNOs from the 3E\_4 simulation ({\it blue}).}
\label{fig:lowincpop_3E}
\end{figure}

Numerous works have proposed that the existence of $\sim$Mars- to $\sim$Earth-mass embryos in the early outer solar system may explain the origin of the population of detached TNOs on large ($a\gtrsim$100 au) orbits \citep{gladchan06, siltre18, huangglad24}. In Figure \ref{fig:lowincpop_3E}, we show an illustrative example of a Kuiper belt formation simulation (3E\_4) that initially includes three separate Earth-mass embryos in the primordial belt. In Panels A and B, we show the temporal evolution of each embryo's pericenter and semimajor axis, respectively. While prior works have suggested that efficiently-detaching embryos occupy low-inclination orbits with $30\lesssim q\lesssim40$ au and $a\gtrsim200$ au, these types of embryo orbits are rarely seen in our simulations. Simulation 3E\_4 is the only one of our six 3E simulations that features an embryo spending a notable amount of time on a semimajor axis over 100 au. In this simulation, Embryo 2 is above 100 au from $t\simeq$100--120 Myrs. However, this is not the embryo that most strongly affects the distribution of TNO orbits. Another embryo (Embryo 1) reaches a semimajor axis of $\sim$48 au after $\sim$100 Myrs. Over the next $\sim$1 Gyr, dynamical friction from its interactions with the TNO population circularizes its orbit, pulling the pericenter from $\sim$40 au to $\sim$48 au. Unlike the other two embryos, this is a stable orbit, and the embryo persists until the end of the simulation.

The presence of the embryos in 3E\_4 has a dramatic impact on the population of detached ($q>40$ au) TNOs. Unlike the Control simulations shown in Figure \ref{fig:lowincpop}, there is now a large population of detached TNOs with semimajor axes between $95<a<400$ au and $i<25^{\circ}$, as can be seen in Figure \ref{fig:lowincpop_3E}C. However, when we compare semimajor axes of the simulated and observed TNOs in Figure \ref{fig:lowincpop_3E}D, it is very clear that simulated low-$i$, high-$a$ detached bodies in 3E\_4 follow a different semimajor axis distribution than the observed bodies. Amongst the observed detached TNOs in this region, the median semimajor axis is $\sim$280 au, but the detached TNOs in the 3E\_4 simulation have a significantly smaller median semimajor axis of $\sim$150 au. Moreover, the 90th percentile of the simulated 3E\_4 detached TNOs semimajor axis distribution is 261 au, while 6 of the 11 observed TNOs have semimajor axes beyond this value. This discrepancy between the observed detached TNOs and the simulated 3E\_4 ones exists {\it despite the fact} that our simulated results do not account for observational bias. Observing bias will skew the simulated sample even more toward lower semimajor axes and further from the observed sample, since the discovery probability of large semimajor axis objects (whose average distance is also large) should scale strongly with their frequency of perihelion passage, or the inverse of their orbital period ($a^{-3/2}$). 

To estimate the strength of this observational bias, we consult the observational survey simulator of \citet{sheptruj16}, as this survey has detected 5 of the 11 members of the distant low-inclination detached population ($q>40$ au; $95<a<400$ au; $i<25^{\circ}$). We use this simulator to estimate the discovery probability of a hypothetical detached population. We assume that this hypothetical detached population has an inclination distribution following $\sin i$ multiplied by a Gaussian centered at 19.1$^{\circ}$ with a width of 6.9$^{\circ}$ \citep{gulb10}. In addition, we assume that semimajor axes are uniformly distributed between 95 and 400 au in the perihelion range of 40--85 au (which brackets the observed perihelion sample). Perihelia are drawn from a uniform distribution between 40 au and $a$. All other orbital elements ($\omega$, $\Omega$, and $M$) are drawn randomly from uniform distributions. Finally, we assume a $\alpha=0.9$ power-law bright-end size frequency distribution (SFD) that transitions to a $\alpha=0.4$ faint-end SFD at $H=7.7$, which is similar to the ``knee'' distribution from \citep{law18} we reference previously.

\begin{figure}
\centering
\includegraphics[scale=0.4]{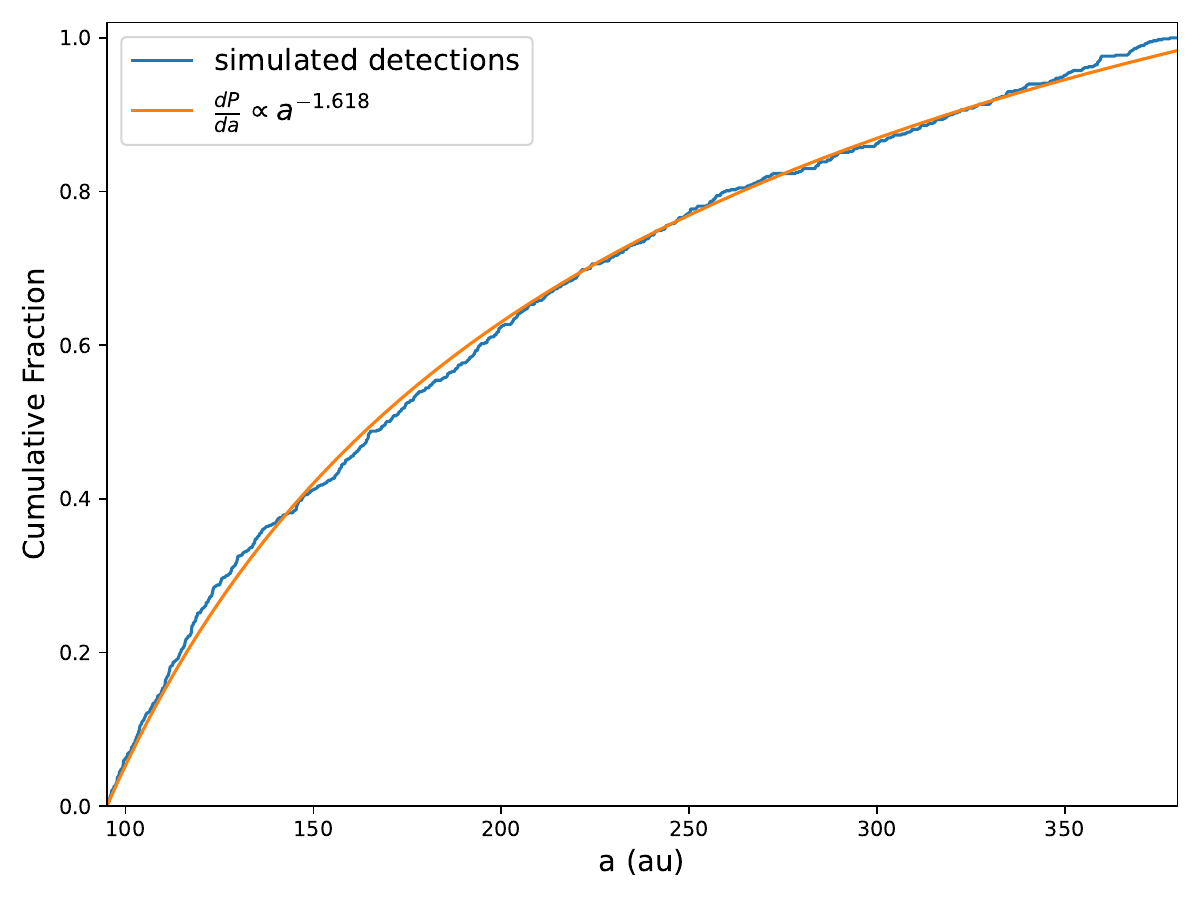}
\caption{Semimajor axis distribution of simulated detections of a hypothetical detached TNO population using the \citet{sheptruj16} survey simulator are shown in blue. Orange shows the best-fit power-law to these simulated detections.}
\label{fig:chadfit}
\end{figure}

Making these assumptions for a hypothetical population, we randomly draw from this population, and we use the \citet{sheptruj16} survey simulator until we simulate 812 detections with inclinations under 25$^{\circ}$ and perihelia between 40--50 au (which encompasses all but one of the 11 objects in Table 3). The semimajor axis distribution of these simulated detections is shown in Figure \ref{fig:chadfit} along with the best-fit power-law to the distribution. As can be seen, the power-law fit is excellent, and the fit index is -1.62, near the expected index of -3/2. This is especially noteworthy because the \citet{sheptruj16}, \citet{shep19} survey is designed to search for exceptionally distant TNOs ($d>50$ au) and does not follow-up on objects with smaller heliocentric distances. Thus, it should be less biased against large semimajor axes than most other surveys, yet detection probability still falls off with nearly $a^{-1.5}$. 

Our best-fit to \citet{sheptruj16} simulated detections confirms that the detection probability of a distant detached TNO depends very strongly its orbital frequency, and the discrepancy between observations and our 3E\_4 simulation is even larger than that implied in Figure \ref{fig:lowincpop_3E}D. To attempt to account for this observing bias, we perform a weighted bootstrap resampling of our 3E\_4 orbits that have $95<a<400$ au, $q>40$ au, and $i<25^{\circ}$, where each orbit from our simulation is given a weight of $a^{-1.5}$ and randomly resampled according to this weighting. With this bootstrapped simulation sample, we then compare its semimajor axis distribution against the observed set of TNOs from the shaded region of Figure \ref{fig:lowincpop}. In our weighted, resampled simulated distribution of orbits, the 90th percentile semimajor axis has dropped from 261 au down to 213 au. Meanwhile, only 3 of the 11 observed TNOs have semimajor axes below 204 au. The probability of drawing less than 4 of 11 orbits with $a<204$ au from the weighted simulated distribution is just $1.25\times10^{-6}$ according to the binomial statistics. For this binomial probability to exceed 1\%, the 90th percentile of our weighted, simulated semimajor axis would have to be 290 au or larger. Thus, we can confidently reject the null hypothesis that the simulated and observed samples have the same underlying distribution.

In Table 2, we list the 90th percentile of the weighted semimajor axis distribution from 95--400 au for orbits with $q>40$ au and $i<25^{\circ}$ for our other simulations. To bolster simulation statistics, we sample our simulation orbits every 100 Myrs for the final Gyr of each simulation, as even with perihelia over 40 au, many orbits are still capable of semimajor axis evolution via weak scattering and/or resonance sticking on $\sim$100-Myr timescales \citep{bann17, volk18, hadtre24}. If our time sampling yields a set of 10 or more orbits, they are resampled with the weighted bootstrapping method used for 3E\_4, and the 90th percentile semimajor axis is calculated. Otherwise, no calculation is performed. 

Table 2 shows that that none of the 3E simulations manage to replicate observed TNOs well. In fact, the previously discussed 3E\_4 simulation produces the most distant semimajor axis distribution, which is still much too close to the Sun. The evolution shown in Figure \ref{fig:lowincpop_3E} is representative of four of our six 3E simulations, as all of their low-inclination detached population are largely confined to semimajor axes below $\sim$210 au. These four runs also all contain a single Earth-mass embryo that survives on a fairly circular orbit between $a\sim$35--50 au until the end of the simulation. Although our simulations do not include an in-situ cold classical Kuiper belt, it is likely that these long-lived embryos would also overexcite this belt's orbits compared to the actual solar system. Meanwhile, the other two simulations (3E\_5 and 3E\_6) lose all embryos within the first 50 Myrs of the simulation, typically via ejection, and they do not contain a significant number of detached TNOs on low-inclination orbits. 

Examining the third column of Table 2, we find that almost none of our runs with sub-Earth-mass embryos place significant numbers of TNOs on orbits with $i<25^{\circ}$, $q>40$ au, and $95<a<400$ au. The large majority of these runs finish with well under 1\% of surviving Kuiper belt objects on such orbits. This would correspond to a trapping efficiency during Kuiper belt formation of $\sim$10$^{-5}$ or lower, and the mass that the observed population implies for the primordial Kuiper belt would be extreme \citep[e.g.][]{trujshep14, beau23}. This strongly suggests that embryos of Mars-mass and below are not efficient at detaching TNO perihelia \citep{gladchan06}, even when 10+ embryos are initially present. There are only six simulations (20M\_3, 20M\_4, 20M\_5, 20M\_6, 5M\_3, and 2M\_3) that possess 10 or more time-sampled orbits with $i<25^{\circ}$, $q>40$ au, and $95<a<400$ au during the final Gyr.  For these six simulations, none provide a good match to the observed population. 5M\_3 has the highest 90th percentile of the weighted bootstrapped semimajor axis distribution for $i<25^{\circ}$, $q>40$ au, and $95<a<400$ au orbits, and it is 155 au, which encompasses only 2 of the 11 observed TNOs in this region. The other 5 runs all feature semimajor axis distributions whose 90th percentile values only encompass 1 or 2 of the 11 observed TNOs in this region. The probability of drawing 9--10 out of 11 objects beyond the 90th percentile semimajor axis is between $\sim$10$^{-10}$--10$^{-7}$ according to binomial statistics. 

\begin{figure}
\centering
\includegraphics[scale=0.4]{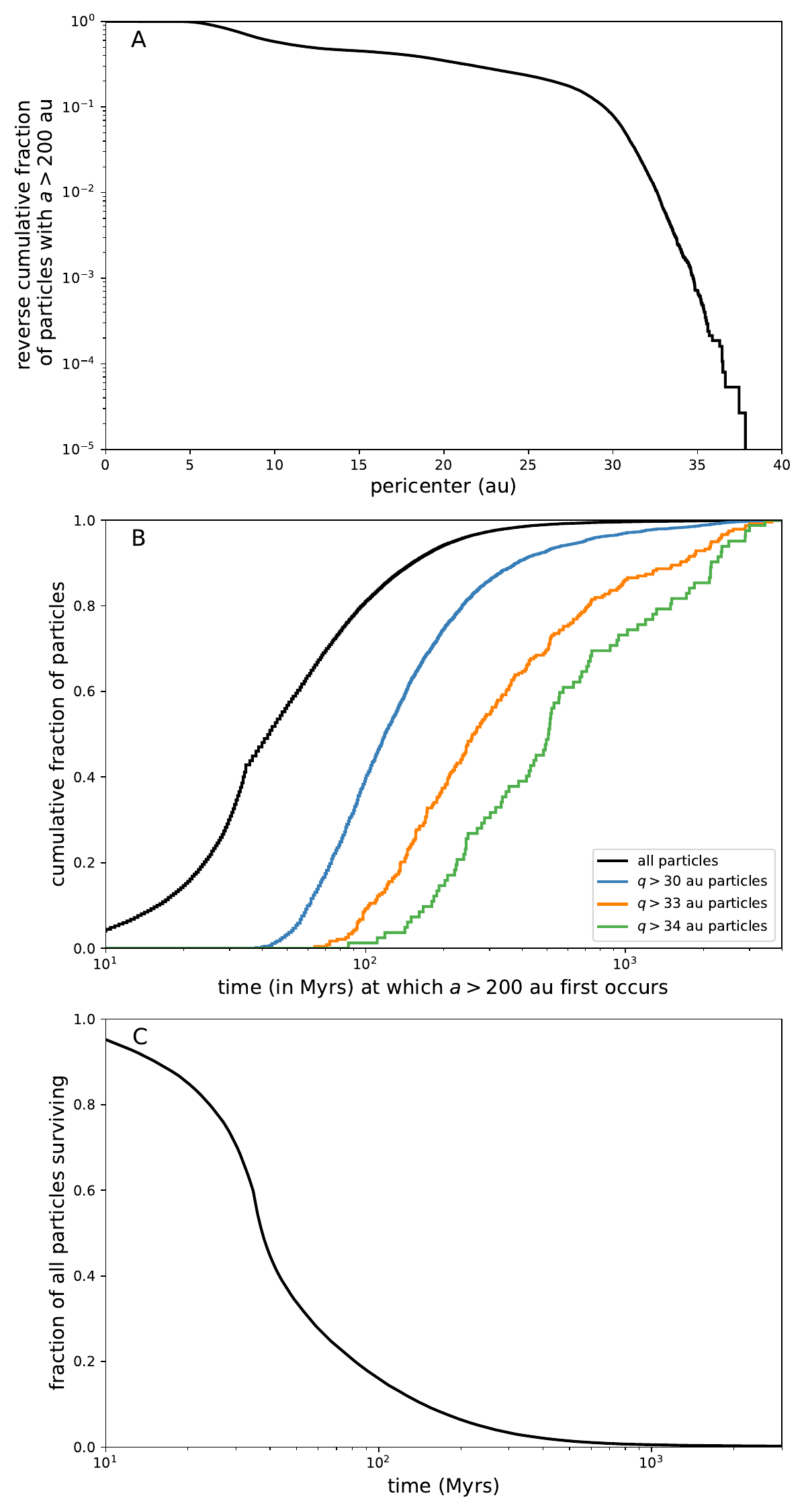}
\caption{{\bf A:} Reverse cumulative distribution of the pericenters of all particles recorded reaching a semimajor axis over 200 au in the 200Pa simulation from \citet{kaib24}. {\bf B:} Cumulative distribution of simulation times at which particles reach semimajor axes over 200 au in the 200Pa simulation from \citet{kaib24}. Distributions are shown for all particles ({\it black}) as well as for particles with pericenter beyond 30 au ({\it blue}), 33 au ({\it orange}), and 34 au ({\it green}). {\bf C:} The fraction of particles still remaining as a function of time in the 200Pa simulation from \citet{kaib24}.}
\label{fig:supprob}
\end{figure}

Among our six sub-Earth-mass embryo simulations that populate orbits with $i<25^{\circ}$, $q>40$ au, and $95<a<400$ au, two (20M\_3 and 20M\_4) feature multiple embryos that persist for over 1 Gyr on low-eccentricity orbits with pericenters between 30--40 au. 2M\_3 is the only one of these systems to feature an embryo on an orbit with $a>100$ au and $q\simeq30$ au for $\sim$100 Myrs or longer, which is similar to the orbital scenarios proposed in \citet{huang22} and \citet{huangglad24}. However, in this simulation, the embryo occupies this orbit from $t=$300--400 Myrs, and it spends the prior 200 Myrs on a semimajor axis fluctuating between 40--100 au. Thus, our Kuiper belt formation simulations including embryos suggest orbital evolution tracks that temporarily deposit embryos into orbits with $a>200$ au and $q$ beyond 30 au are quite improbable. However, the small number of embryos our simulations possess make it difficult to precisely estimate this probability. To access much greater numbers of particles, we can instead examine the orbital evolution of the `200Pa' non-embryo Kuiper belt simulation from \citet{kaib24}. This simulation contains $\sim$10$^5$ Pluto-mass and sub-Pluto-mass particles that are scattered outward by the giant planets as they migrate. This particular simulation is chosen because it broadly reproduces basic features of the Kuiper belt and giant planets' orbits \citep{kaib24}. Because the 200Pa particles have smaller masses, dynamical friction will not play a notable role in their orbital evolution. This is in contrast to the behavior in Figures \ref{fig:lowincpop_3E}A and \ref{fig:lowincpop_3E}B that demonstrates Earth-mass embryos can be influenced by dynamical friction. However, judging from Figure \ref{fig:lowincpop}, the timescale of this influence seems to be $\gtrsim$100 Myrs, typically longer than planetary scattering timescales. 

In Figure \ref{fig:supprob}A, we plot the pericenter distribution of every particle in the \citet{kaib24} 200Pa simulation that reaches a semimajor axis over 200 au. (The pericenters are recorded after an orbit switches from $a<200$ au to $a>200$ au.) We see that only 7--8\% of particles attaining $a>200$ au orbits do so at pericenters beyond 30 au, and the fraction becomes vanishingly small ($<10^{-3}$) if we restrict ourselves to pericenters beyond $q>35$ au. Moreover, this fraction continues to fall steeply with increasing pericenter, and it is below $10^{-5}$ at $q\simeq40$ au. In addition, the planetary scattering that drives orbits to $a>200$ au at $q>30$ au is weak, and such scattering typically proceeds over 100 Myrs or longer \citep{fern81, dqt87}. This is confirmed in Figure \ref{fig:supprob}B where we look at the distribution of times that it takes our sample of particles to attain $a>200$ au orbits. For the overall sample, some particles can reach this semimajor axis in under 10 Myrs, and the median time is $\sim$40 Myrs. However, this median timescale increases to 100 Myrs when we restrict ourselves to pericenters over 30 au, and it increases to $\sim$500 Myrs if we only consider pericenters beyond 34 au. For orbits with pericenter over 34 au, we have nearly no examples of particles reaching $a>200$ au in under 100 Myrs. This presents a major problem for an embryo origin to the detached TNO population because the supply of TNOs whose pericenters can be detached falls quickly with time during Kuiper belt formation. In Figure \ref{fig:supprob}C, we plot the fraction of initial particles that still remain in the 200Pa simulation as a function of time. At $t=100$ Myrs, it has fallen to $\sim$15\%, and at $t=500$ Myrs, it is 1--2\%. Thus, if an embryo is able to attain an orbit with $a>200$ au and $q\simeq35$ au, the time it takes to do so greatly diminishes its efficiency of detaching TNOs.

Figure \ref{fig:supprob} strongly suggests that orbital evolution trajectories that deposit objects into orbits with $a>200$ au and $q$ significantly beyond 30 au are quite rare, and when such evolution does occur it typically takes at least $\sim$100 Myrs for bodies to reach this portion of orbital space, at which point the original primordial Kuiper belt population has been depleted by nearly an order of magnitude. However, the  simulations we analyze up to this point are not without shortcomings that can introduce some uncertainty into this conclusion. First, even with the GPU-accelerated feature of GENGA, the 200Pa simulation still contains an order of magnitude less bodies than embarrassingly parallel (CPU-based) simulations of Kuiper belt formation relying on massless test particles \citep[e.g.][]{nes15a, kaibshep16}. Moreover, the evolution of the giant planets in our GENGA simulations do not yield a perfect match to our observed planets' orbits. In contrast, the planetary orbits within simulations using massless test particles are more precisely manipulated with prescribed forces to yield a very close match to the modern giant planet orbits. 

\begin{figure}
\centering
\includegraphics[scale=0.4]{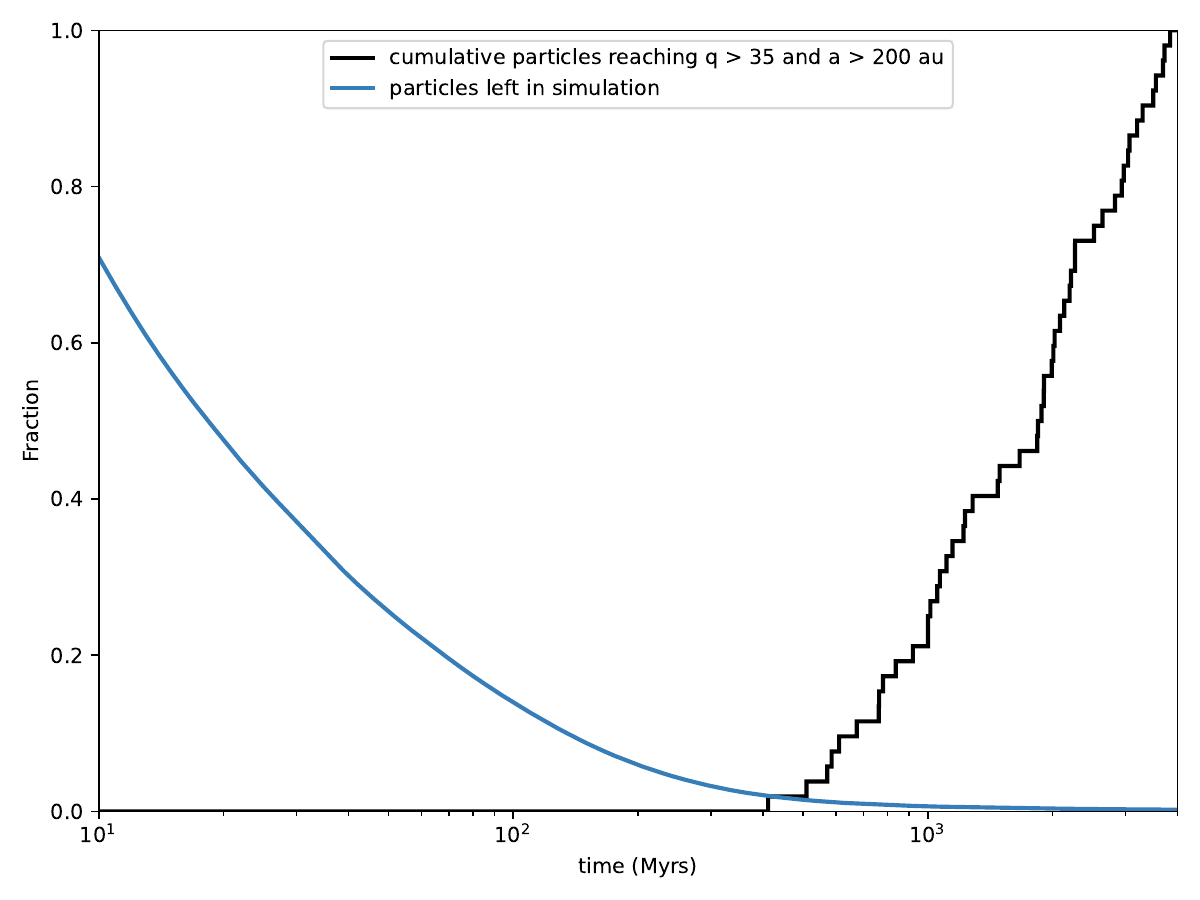}
\caption{The cumulative fraction of particles reaching $q>35$ au and $a>200$ au as a function of time is shown in black for the ``grainy slow'' simulation from \citet{kaibshep16}. The blue curve shows the total fraction of simulation particles remaining as a function of time.}
\label{fig:probGS}
\end{figure}

Thus, we next use the ``grainy slow'' simulation from \citet{kaibshep16} as an additional way to estimate the probability and timing of particles reaching $a>200$ au near $q\sim$30--40 au. As with our self-gravitating GENGA simulations, we find that such evolution is exceedingly rare and occurs well after the start of the simulation. Out of the $10^6$ particles with which the grainy slow simulation begins, only 52 (or 0.0052\%) reach $a>200$ au beyond $q>35$ au. The cumulative fraction of the particles that attain $a>200$ au and $q>35$ au is shown as a function of time in Figure \ref{fig:probGS}. Here we see that no particles reach this region before $t\sim400$ Myrs, and these particles reach this region of orbital space at a median time of well over 1 Gyr. Moreover, the trends of diminishing probability and late, slow evolution continue with increasing pericenter beyond 35 au and towards 40 au. If we increase our pericenter requirement to $q>38$ au, only two particles out of our one million reach $a>200$ au. Furthermore, they do so very late (611 Myrs and 1.219 Gyrs) in the simulation's evolution. At these late times, $\gtrsim$99\% of primordial Kuiper belt particles have been ejected, which means very few particles remain that can also be pulled into a detached orbit by any massive embryo that happens to reach a large semimajor axis, large pericenter orbit. 

None of our Kuiper belt formation simulations including Earth-mass embryos both place an embryo on an orbit with $a>200$ au and $q>35$ au and also avoid having an embryo occupy a closer, nearly circular orbit for a longer period of time (an outcome that existing TNO surveys likely prohibit given such an object's brightness \citep{brown15, shep16}) . This prevents us from understanding the population of detached TNOs generated during the low-probability, previously proposed scenario that a single embryo sculpts the detached TNOs from a high pericenter, high semimajor axis orbit. To address this shortcoming, we again turn to the \citet{kaib24} 200Pa simulation to manually force such an event. This simulation features five particles that attain orbits with $a>200$ au and $q>34$ au orbits within the first 150 Myrs. For each of these five particles, we perform a new additional unique simulation. For each of these simulations, the first stage, where the particle of interest has a semimajor axis below 200 au, is identical to the original 200Pa simulation. However, after the particle of interest attains $a>200$ au, its mass is manually promoted to 1 $M_{\oplus}$, and the 200Pa simulation is reintegrated from that point onward to $t=4$ Gyrs. 

Of the five reintegrations that we performed, two failed to capture any detached TNOs into the shaded region of Figure \ref{fig:lowincpop}, and only one managed to capture more than a single particle into this region. The evolution of the embryo of this particular simulation is shown in Figure \ref{fig:origvspromote}. After being promoted to an Earth-mass embryo at $t=145$ Myrs, we see in Panel B that this embryo maintains a semimajor axis of over 200 au for the next $\sim$35 Myrs before it falls to 100--200 au for the next 300 Myrs. Driven by weak planetary scattering, it then begins a diffusion toward larger semimajor axes (200--1000 au) before ultimately being flagged for ejection after $\sim$1.35 Gyrs. (It should be noted that the particle ejection criterion of these reintegrations is increased from $r=1000$ au to $r=2000$ au to guard against artificially clipping the high end of the semimajor axis distribution of our detached TNOs, although there is no indication this is an issue in our other runs, as discussed in Section \ref{sec:eject}.) During the entirety of this semimajor axis evolution, the pericenter (see Panel A) remains anchored near $q\simeq35$ au. 

It is important to note that our manipulated reintegration of the 200Pa simulation still treats our Earth-mass embryo as a sub-Pluto-mass body in the $t<145$ Myrs portion of the simulation. This initial phase includes a 70-Myr period where this particle occupies an orbit with semimajor axis near 45 au and pericenter near 30--35 au. This is reminiscent of the orbit of the surviving embryo in Figure \ref{fig:lowincpop_3E}'s simulation, which generated a very large population of low-inclination detached bodies with $a\sim100$ au. Although our reintegration does not include an embryo during this early phase, it is quite plausible that such an inclusion would enhance the production of low-inclination detached bodies at small semimajor axes, and the results of our reintegration should be viewed as a best-case scenario for the generation of low-inclination detached bodies on large semimajor axes without producing too many on smaller semimajor axes.

\begin{figure}
\centering
\includegraphics[scale=0.4]{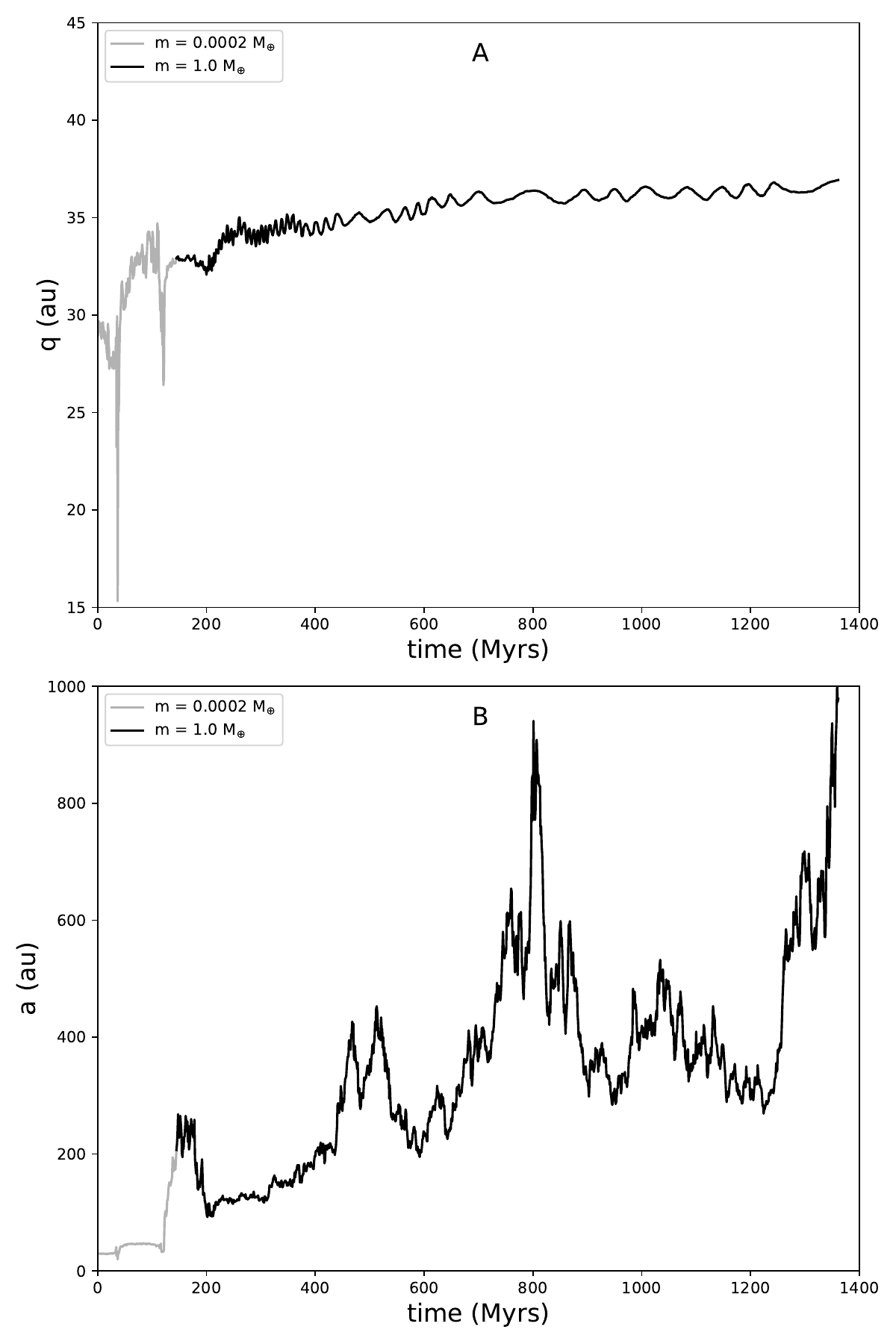}
\caption{{\bf A:} Particle pericenter vs time of a single particle in the 200Pa simulation from \citet{kaib24}. The particle begins the simulation with a mass of 0.0002 M$_{\oplus}$ ({\it gray}) but is promoted to a mass of 1 M$_{\oplus}$ after 145 Myrs ({\it black}). {\bf B:} Semimajor axis vs time of the same particle. This particle is ultimately ejected ($r>2000$ au). All orbital elements are barycentric. }
\label{fig:origvspromote}
\end{figure}

The final orbital distribution of surviving particles from this simulation is shown in Figure \ref{fig:origvspromote}. It is compared with the final distribution from the original 200Pa, and the effects of the Earth-mass embryo are readily apparent. In Panel A, we see that the reintegration finishes with a handful of particles whose pericenters extend out to $\sim$80 au, whereas the pericenter distribution of the original 200Pa stops near $\sim$55 au. When we restrict ourselves to only particles with $q>40$ au in Panel B, we see that the original 200Pa simulation contains a complete absence of particles with $i<25^{\circ}$ and $a>50$ au. In the reintegration, the Earth-mass embryo generates a population of such particles, but it falls off quickly with increasing semimajor axis. When we sample the shaded region of Figure \ref{fig:origvspromote}B every 100 Myrs for the simulation's final Gyr, we compile a sample of 31 orbits. Resampling these orbits with our weighted bootstrap method, we find that the 90th percentile of the semimajor axis distribution is just 136 au, smaller than 9 of the 11 observed TNOs in this region. Thus, we once again find the detached TNOs of our simulation are a poor match to observations. We finally also note that the two other particles trapped in the $95<a<400$ au, $q>40$ au, $i<25^{\circ}$ space in our two other reintegrations had semimajor axes of 126 and 166 au. These are again smaller than over 80\% of the observed detached TNOs in this region of orbital space and again consistent with the idea that a distant embryo will detach too few bodies at large semimajor axes to explain the observed TNO population. 

\begin{figure}
\centering
\includegraphics[scale=0.4]{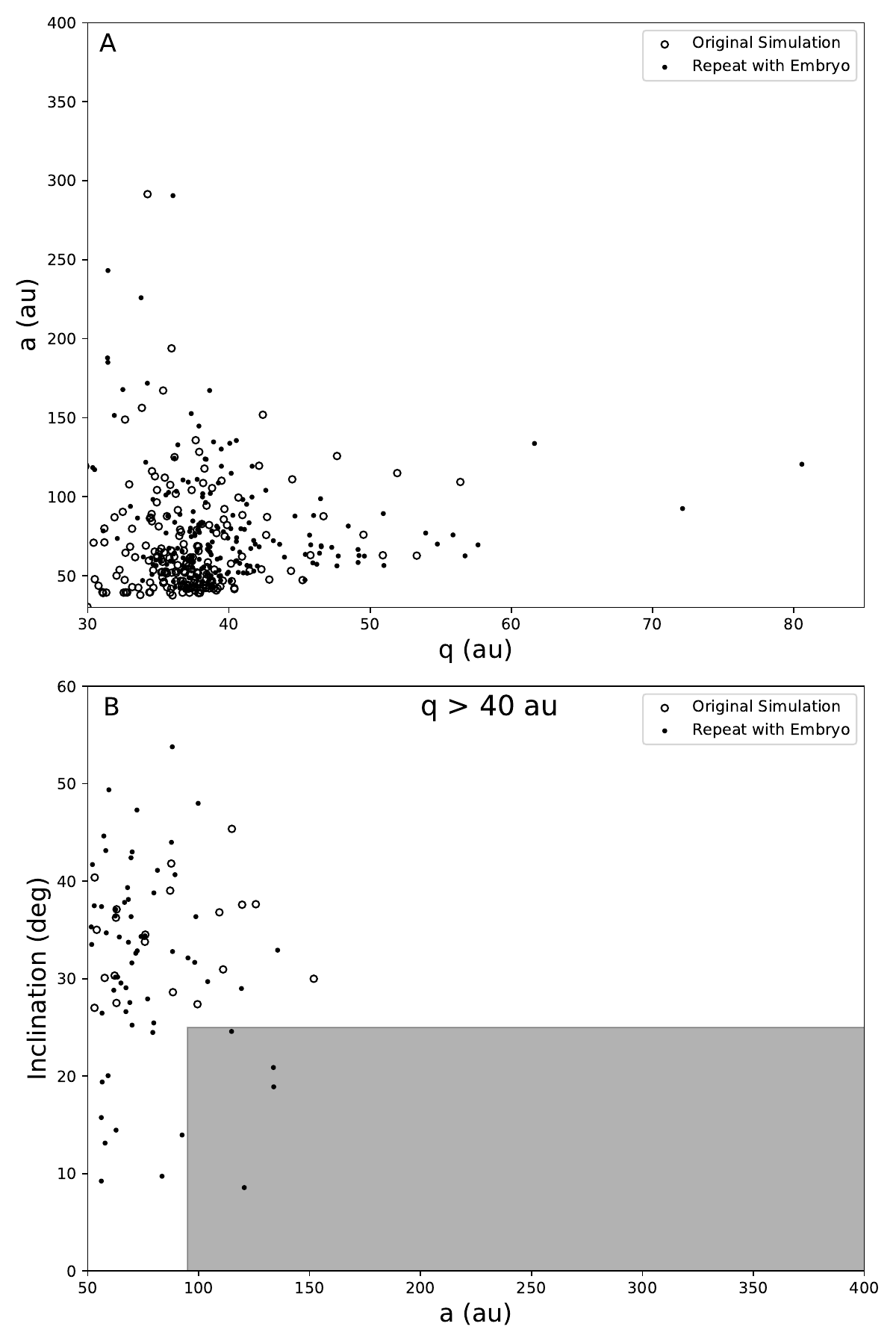}
\caption{{\bf A:} Particle semimajor axes vs pericenters from the 200Pa  simulation from \citet{kaib24} ({\it open symbols}) and a repeat of the simulation ({\it solid symbols}) in which one selected particle's mass was promoted to 1 M$_{\oplus}$.  {\bf B:} Particle inclinations vs semimajor axes from the same two simulations from Panel A. Only particles with pericenter over 40 au are shown. The shaded area marks the region of orbital space that is difficult to populate in our control simulations (see Figure 1).}
\label{fig:origvspromote}
\end{figure}

\section{Summary \& Conclusions}\label{sec:con}

The origins of the detached population of TNOs remain a dynamical puzzle. This is especially true for detached TNOs whose inclinations are below 25$^{\circ}$ and cannot therefore be easily explained with the Kozai mechanism. We find that although planetary mass embryos embedded in the primordial Kuiper belt can generate a population of low-inclination TNOs with perihelia detached from the giant planets, this population is statistically incompatible with the present-day TNO orbital distribution. There are two main reasons to disfavor an embryo scenario for detached TNO population:

\begin{enumerate}

\item{The probability and length of time for an embryo to reach the large semimajor axis ($a\gtrsim200$ au), high perihelion ($q\simeq$30--40 au) orbit proposed in prior works to detach TNO perihelia is inconsistent with Kuiper belt formation dynamics. Using sub-Pluto bodies and test particles as tracers of dynamical pathways to $a>200$ au, we find there is less a 0.1\% probability that if an embryo's orbit exceeds $a>200$ au, it will do so beyond $q>35$ au. Second, if an embryo does manage to exceed $a>200$ beyond $q\gtrsim35$ au, it will require at least a few hundred Myrs of dynamical evolution to do so, since planetary scattering is weak at such pericenter values. After hundreds of Myrs of evolution, the primordial Kuiper belt's population will be depleted by 90-99\% of its initial value, meaning that the source population for detached TNOs will be much smaller than prior work assumes. }

\item{Disregarding the probability and time necessary for an embryo to attain the desired `detaching' orbit, we still find reason to disfavor this origin for the detached TNO population. The reason is that there is compelling observational evidence that the population of low-inclination detached TNOs with semimajor axes between $\sim$100--400 au is heavily weighted toward the large semimajor axis portion of this range. This abundance of large-semimajor-axis TNOs exists in spite of the strong observational bias we expect against the discovery of real detached TNOs on the largest semimajor axes. Meanwhile, our simulations that include embryos detach many more TNOs at the smaller end of this range than the larger end. None of our simulations that include embryos are able to replicate the abundance of large semimajor axis orbits seen in the real detached population. This is the case even when we manually force an embryo onto a $q\gtrsim35$ au, $a\gtrsim200$ au orbit during the dispersal of the primordial Kuiper belt. }

\end{enumerate}

While the existence of Mars--Earth-mass bodies in the primordial Kuiper belt is plausible, we conclude that they are unlikely to be the primary means through which TNO perihelia have been detached from the giant planets' dynamical influence. It is also not clear whether a distant, undetected planet or an early encounter with a stellar (or substellar) body can replicate the observed features of the detached TNO orbital distribution that we highlight in this work. These alternative mechanisms should be evaluated in a similar manner to assess their viability as the origin mechanism of detached TNOs. 

\section{Acknowledgements}

NAK's contributions were supported from NSF CAREER award 2405121 and NASA Emerging Worlds grant 80NSSC23K0771. CAT's contributions were made possible in part by the State of Arizona Technology \& Research Initiative Fund. This work used the Expanse GPU cluster at the San Diego Computing Center through allocation AST190052 from the Advanced Cyberstructure Coordination Ecosystem: Services \& Support (ACCESS) program, which is supported by National Science Foundation grants \#2138259, \#2138286, \#2138307, \#2137603, and \#2138296. We thank reviewer Kathryn Volk as well as another anonymous reviewer whose comments and suggestions greatly improved this work. \clearpage

\bibliographystyle{apj}
\bibliography{SuperPlutos}

\end{document}